\documentclass[9pt,twocolumn,twoside]{optica_arxiv}
\setboolean{shortarticle}{false}
\setboolean{minireview}{false}
\setboolean{displaycopyright}{false}

\usepackage{makecell} 
\usepackage{stfloats}

\makeatletter

\newcommand{\Rmnum}[1]{\expandafter\@slowromancap\romannumeral #1@}
\makeatother

\title{Single-Photon Fourier Transform}

\author[1,2,3,4]{Zhen Yang}
\author[1,2,4]{Zeng-Quan Yan}
\author[1,2]{Li Wang}
\author[1,2]{Xiao-Wei Wang}
\author[3,5]{Ka-Di Zhu}
\author[1,2,*]{Xian-Min Jin}

\affil[1]{Center for Integrated Quantum Information Technologies (IQIT), School of Physics and Astronomy and State Key Laboratory of Photonics and Communications, Shanghai Jiao Tong University, Shanghai, China.}
\affil[2]{Hefei National Laboratory, Hefei, China.}
\affil[3]{Key Laboratory of Artificial Structures and Quantum Control (Ministry of Education), School of Physics and Astronomy, Shanghai Jiao Tong University, 800 DongChuan Road, Shanghai 200240, China.}
\affil[4]{These authors contributed equally to this work.}
\affil[5]{zhukadi@sjtu.edu.cn}
\affil[*]{Corresponding author:  xianmin.jin@sjtu.edu.cn}

\begin{abstract}    
The extraction of information carried by light plays an increasingly important role in optical communication, imaging, and detection. However, the information can only be successfully extracted when the light pulse is comparably strong, leaving untouched scenarios where survived photons are extremely sparse. Here, we propose and experimentally demonstrate a single-photon Fourier transform scheme. By retrieving the implicit correlation shared in the sparse single-photon stream globally, we are able to precisely classify each photon and synchronously extract multiple ultra-weak signals with high fidelity against extreme environments. Our experiment results give a full picture of the scheme in terms of multi-terminal expandability, wide frequency adaptability, 125 dB loss tolerance, and -10.4 dB signal-to-noise ratio robustness. Even when the pulse repetition frequencies of all terminals are the same, we can still recognize the free-running clock drift and separate different messages. Our work can be a general scheme to extend the capability boundary for all the extremely low-light-flux scenarios, and makes many challenging tasks possible, such as in-orbit optical communication network with complex topology, navigation in extremely lossy and noisy environments, and wide-range single-photon imaging with multi-source illumination.
\end{abstract}

\begin{document}
\maketitle
	
\section{Introduction}
\begin{figure*}[t]
	\centering
	\includegraphics[width=1.9\columnwidth]{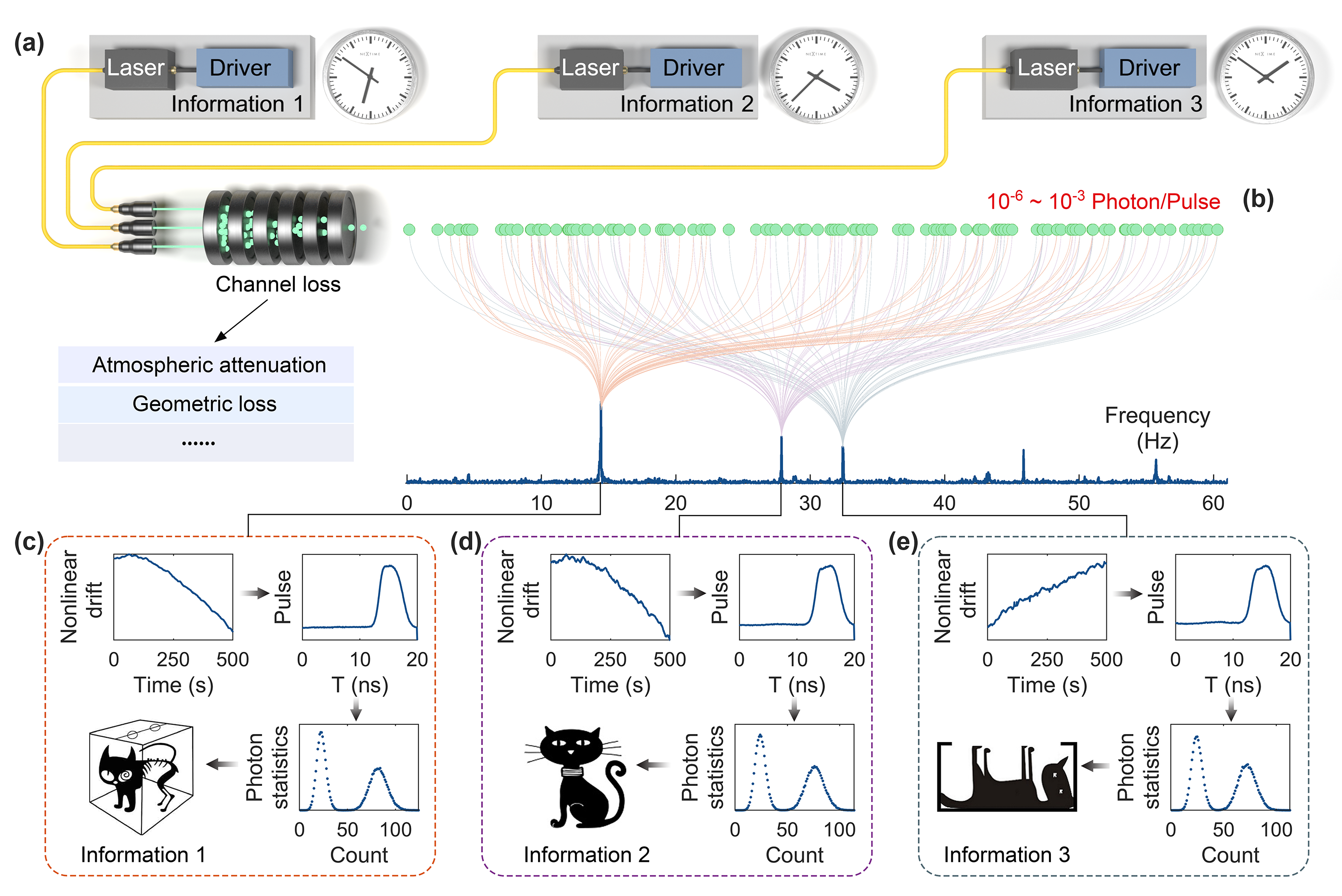}\\
	\caption{\textbf{Illustration of the single-photon Fourier transform (SPFT) scheme.}
		\textbf{(a)} Schematic of the experimental set-up. Three 520 nm laser diodes (LDs) are all modulated to 50 MHz using the return-to-zero on-off keying (RZ-OOK) method and cyclically send three binary sequences encoding different pictures, respectively. We label the three signals as information 1, 2, 3. \textbf{(b)} By performing SPFT scheme on the mixed sparse single-photon stream, we can extract the frequency characteristics and classify photons to correct terminals. The marked peaks in the frequency domain waveform correspond to the linear clock drifts of three transmitting terminals, respectively. \textbf{(c)}-\textbf{(e)} Based on the previous step, we further calculate the nonlinear clock drift of each transmitting terminal, realizing the high precision synchronization and restoring pulse waveform. Then, we utilize photon statistics to separate signal from noise and achieve the retrieval of binary sequences. Finally, the transmitted picture information can be reconstructed by applying the decoding procedure.}
	    \label{fig1}
\end{figure*}

Light has both fundamental interest and wide applications in diverse fields like optical communication \cite{fiber2001optical, FSO2016, kaushal2016optical}, imaging \cite{laserimaging, opticalimaging}, and detection \cite{laserranging, laserranging2018}. In some substantial scenarios such as long-distance optical communication \cite{LLCD, underwater, zhang2021high}, quantum communication \cite{quantum2007, timesecure}, and low-light imaging \cite{First-PhotonImaging, 2020singlephotonimaging}, the detectable photons are extremely sparse owing to the high channel loss or the security issue. Making the information extraction from the sparse single-photon stream a foundation stone for many valuable applications.

For instance, deep-space or underwater optical communications need to extract information from survived photons that fade far below one photon per pulse after huge geometric loss and channel loss \cite{2011dsoc, yan2021PICOC, underwaterPICOC}. Quantum communication \cite{BB84} needs to manipulate the phase \cite{2012phaseQKD, timebin2013, phase2018, chen2020TFphase} or polarization \cite{PolarizationEncoding, PolarizationEncoding2, PolarizationEncoding3, PolarizationEncoding4siliconphotonics} of photons to share secure keys. After experiencing the channel loss, only about $10^{-4}$ photons per pulse survived at receiving terminal. Single-photon imaging needs to reconstruct images by collecting reflected photons that are less than one photon per pixel to overcome harsh detection environments \cite{image2016photon, li2021single, LiThresholded}. The introduction of a single-photon detector (SPD) gives the above systems the ability to detect weak signals \cite{SPD, 2015SPD, chipSPD}. However, widely researched technologies of information extraction combined with SPDs are still restricted to interacting with only one light source. To meet the needs of universal applications, it is essential to develop a more versatile approach that can distinguish multiple different signals from the mixed sparse photon stream respectively. 

Here, we propose a single-photon Fourier transform (SPFT) scheme that exploits the implicit correlation shared in photon stream to separate mixed weak signals. Concretely, by identifying the slight difference in clock drift patterns, we can synchronously extract multiple independent information from the mixed sparse single-photon stream where only $10^{-6}$ photons per pulse and contributed by serval time-individual sources [as shown in Fig. \ref{fig1}]. Based on a system with three transmitting terminals and one receiving terminal equipped with a single-pixel SPD, we experimentally demonstrate the key performance of our scheme. The results show not only the excellent multi-terminal expandability, but also the wide frequency adaptability, the strong robustness against ultra-high channel loss and low signal-to-noise ratio (SNR) environments. 

\section{Principle of SPFT Scheme}
\begin{figure*}[t]
	\centering
	\includegraphics[width=1.8\columnwidth]{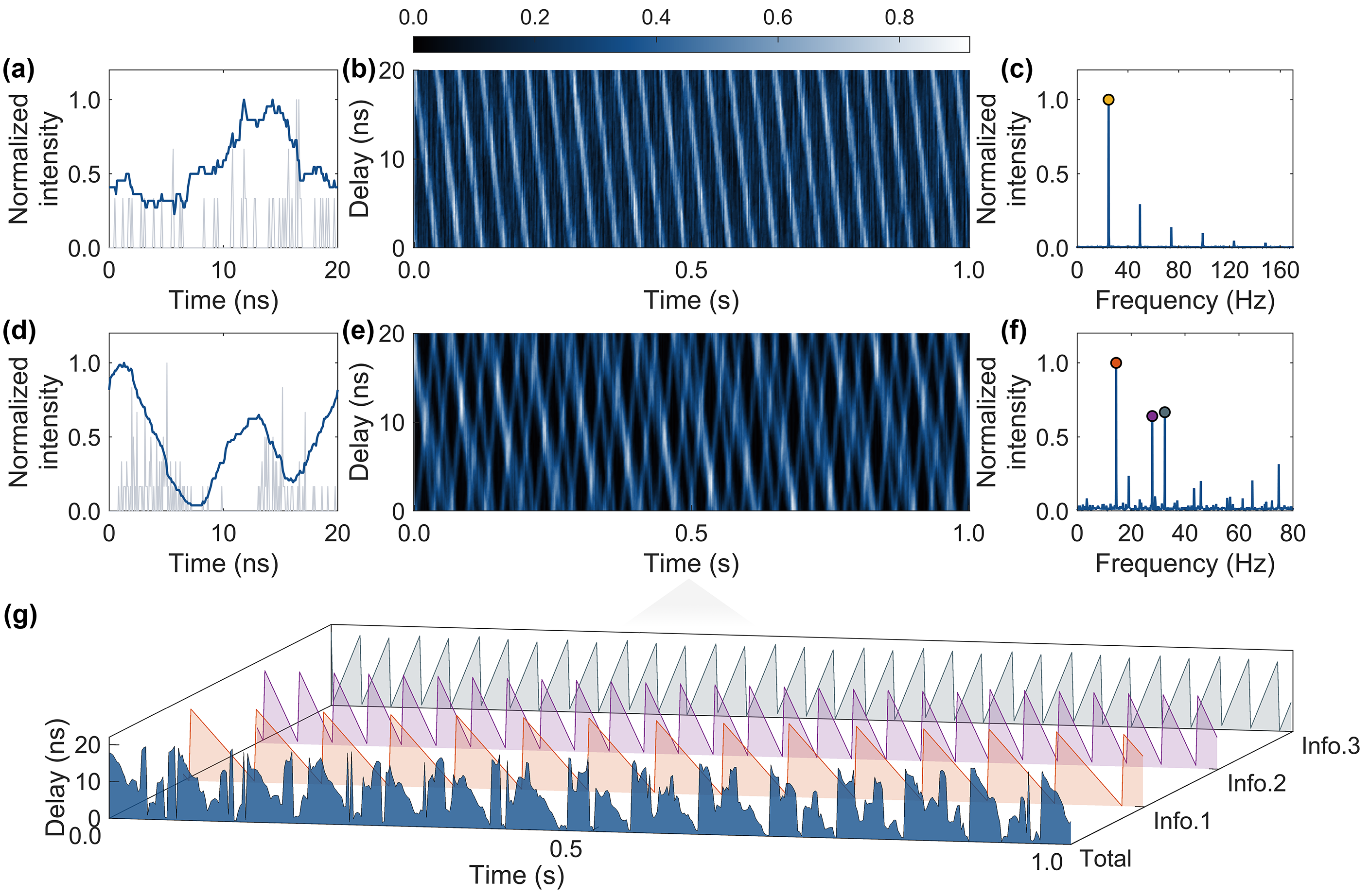}
	\caption{\textbf{Retrieval of linear drift rates among transceiver clocks via SPFT scheme.}
		\textbf{(a)} The coincidence of single-signal photons in a time slot. The gray curve is the delay of photons relative to the pulse period. The dark blue curve is the coincidence waveform of photons. Here, the time slot width $\tau_{slot}=2$ ms, the pulse period $T_{0}=20$ ns, the gate width $\tau_{gate}=5$ ns. \textbf{(b)} The periodic oscillation of coincidence delay with time during one-second length. \textbf{(c)} The frequency domain waveform of the coincidence delay obtained by applying Fourier transform. \textbf{(d)} The coincidence of photon stream containing three individual signals in a 2 ms time slot. The gray curve is the delay of mixed photons relative to the pulse period 20 ns. The dark blue curve is the mixed coincidence waveform with a 5 ns gate width superimposed by several individual pulses. \textbf{(e)} The coincidence delay oscillation of mixed photons with time, which actually includes multiple independent periodic oscillations. \textbf{(f)} The frequency domain waveform of the intricate coincidence delay. \textbf{(g)} The dark blue curve is the experimental result about superimposed coincidence delay oscillation of mixed signals. The rest curves are the theoretical independent oscillations of three signals, respectively.} 
	\label{fig2}
\end{figure*}

The SPFT scheme we proposed is based on the fact that each source and detector have an independent reference clock, and the system adopts the pulse modulation strategy. Since the oscillations of any two remote clocks are inevitably affected by internal or environmental factors \cite{kaushal2016optical, frequencyoffset, TimeDeepspace}, there is always a relative drift between transmitter and receiver clocks. 

We first discuss the case when the sparse single-photon stream is contributed by one transmitting terminal. In extracting the correlation among a few photons, we divide the arrival time of the photon stream uniformly into $\tau_{slot}$ ms width time slots. Then we calculate the remainders of photon arrival time $t_a$ upon division by pulse period $T_0$ in each time slot: $t_{mod}=t_a\ mod\ T_0$. Due to the extremely sparse photons in the time slot, it is hard to get valid information by directly observing the calculated result [as shown by the gray curve in Fig. \ref{fig2}(a)]. Here, we introduce the concept of photon coincidence. The coincidence event, in quantum optics, refers to the simultaneous response of a few photons below a given gate width which usually ranges from several picoseconds to nanoseconds, reflecting the temporal correlation of photons. The calculated $t_{mod}$ allows us to trace photons far apart back into a pulse period. By further applying the coincidence concept with a gate width of $\tau_{gate}$ ns (see Supplementary for the selections of $\tau_{slot}$ and $\tau_{gate}$), we can obtain the coincidence waveform shown by the dark blue curve in Fig. \ref{fig2}(a). Afterwards, we extract the value corresponding to the coincident peak as the delay for each time slot and arrange them together in chronological order. Fig. \ref{fig2}(b) shows the magnified one-second data, we notice that the fluctuation of coincidence delay with time has a certain periodicity, which can be described as a sawtooth wave and is related to the linear drift rate between clocks (specific derivation can be found in Section \ref{Methods}). 

More generally, considering a mixed sparse single-photon stream contributed by several transmitting terminals with independent reference clocks. In this case, the retrieved waveform of each time slot [Fig. \ref{fig2}(d)] presents the mixture of several individual pulses. Due to the oscillating drift difference between clocks, the relative pulse position in each time slot keeps drifting, resulting in the overall coincidence delay in Fig. \ref{fig2}(e) being far more complex than in Fig. \ref{fig2}(b). Despite the intricate overall coincidence delay, we can still precisely extract the linear clock drift of each transmitting terminal from the mixed sparse single-photon stream via the Fourier transform.

We theoretically decompose the superposed coincidence delay into several independent parts corresponding to each terminal [Fig. \ref{fig2}(g)], which can be expressed as

\begin{equation}
\begin{array}{lc}
\begin{aligned}
x^{\prime}\left(t\right)&=\sum\limits_{j=1}^{s}\sum\limits_{n=1}^{+\infty}\frac{p_j T_0}{\pi n}\left(-1\right)^{n+1}\sin\left(2\pi n \frac{k_j}{T_0} t+\varphi_j\right)+\frac{T_0}{2},\\
\end{aligned}
\end{array}
\end{equation}
where $j$ indicates different signals and the total number of received signals is $s$. In our demonstrated experiments, we choose $s=3$. $p_j$ means the proportion of $j$-signal in the mixture, which satisfies $\sum\limits_{j=1}^{s}p_j=1$. $k_j$ is the linear drift rate between the receiver and the $j$-signal transmitter clock. $\varphi_j$ represents the initial oscillation phase of the coincidence delay belonging to $j$-signal.

Then we utilize the global analysis feature of Fourier transform to correlate each partially associated time slots, which can characterize all the periodic behaviors in the time domain. And the corresponding frequency domain representation indicating the clock drifts of mixed signals can be theoretically expressed as 

\begin{equation}
\begin{array}{lc}
\begin{aligned}

X^{\prime}\left(m f^{\prime}\right)=&\frac{1}{T^{\prime}}\int_{0}^{T^{\prime}}x^{\prime}\left(t\right)\exp{\left(-{\rm {\mathbf i}} 2\pi m f^{\prime} t\right)}\mathrm{d}t\\
=&\sum\limits_{j=1}^{s}\sum\limits_{n=1}^{+\infty}\frac{{\rm {\mathbf i}} p_j T_0}{2\pi n}\left(-1\right)^{n}\left[\exp{\left({\rm {\mathbf i}} \varphi_j\right)}\delta_{n \frac{k_j}{T_0},m f^{\prime}}\right.\\
&-\left.\exp{\left(-{\rm {\mathbf i}} \varphi_j\right)}\delta_{n \frac{k_j}{T_0},-m f^{\prime}}\right],\\

\end{aligned}
\end{array}
\end{equation}
where $T^{\prime}=1/f^{\prime}$, $f^{\prime}$ denotes the fundamental frequency of Fourier expansion of the superposed sawtooth wave $x^{\prime}\left(t\right)$. Term $\delta_{n \frac{k_j}{T_0},m f^{\prime}}$ is the Kronecker delta function. Fig. \ref{fig2}(f) shows the experimental waveform of the frequency domain. The marked characteristic peaks are mapped to three signals and can be used to calculate the practical linear part of clock drifts. With the obtained linear clock drift rate $k_j$, we can further classify photons in the mixed sparse single-photon stream and establish the coarse synchronization procedure between the receiving terminal and each transmitting terminal [see Fig. \ref{fig1}(b) and Supplementary for details]. 

Based on the linear clock drift calibrating, we can further trace the high order nonlinear clock drift of each transmitting terminal, realizing the fine time synchronization of each signal in the single-photon stream [see retrieved pulse waveforms in Figs. \ref{fig1}(c)-\ref{fig1}(e) and Supplementary for details]. Afterwards, we perform the time evolution feature of photon statistics to separate signal and noise [see photon statistics in Figs. \ref{fig1}(c)-\ref{fig1}(e) and Supplementary for details]. Finally, all information can be reconstructed with a bit error rate (BER) well beneath the forward error correction (FEC) threshold of $3.8 \times 10^{-3}$ \cite{FEC2016, FEC2018} under an extremely low signal level. BER lower than the FEC threshold means the ability to correct all bit errors, representing the successful establishment of synchronously extracting multiple time-individual information from the mixed sparse single-photon stream via our scheme [Figs. \ref{fig1}(c)-\ref{fig1}(e)]. 

In a practical system, the mixed single-photon stream could be extremely sparse due to the channel loss, causing errors or missing slots in the coincidence delay oscillation. Nevertheless, the global analysis feature of our scheme can retrieve the characteristics of different signals regardless of these errors.

\section{Experimental results}

To show the robustness of our SPFT scheme, we construct a many-to-one experimental system composed of three transmitting terminals and one receiving terminal. Each terminal is driven by an independent reference clock, causing the asynchronous time reference in the network. At transmitting terminals, we equip three 520 nm laser diodes (LD) to generate light pulses. By encoding three pictures with specific pulse repetition frequencies and using two optical circulators to transmit, we can obtain the mixed photon stream [Fig. \ref{fig1}(a)]. At the receiving terminal, a single-pixel SPD and a time-to-digital converter (TDC) are employed to detect the arrival time of the mixed photon stream (see Supplementary for details). Then we applied our SPFT scheme to process the collected data, experimentally exploring its frequency adaptability, loss tolerance, and practicality.

\subsection*{A. Wide frequency adaptability of SPFT scheme}

\begin{figure*}[pt]
	\centering
	\includegraphics[width=1.7\columnwidth]{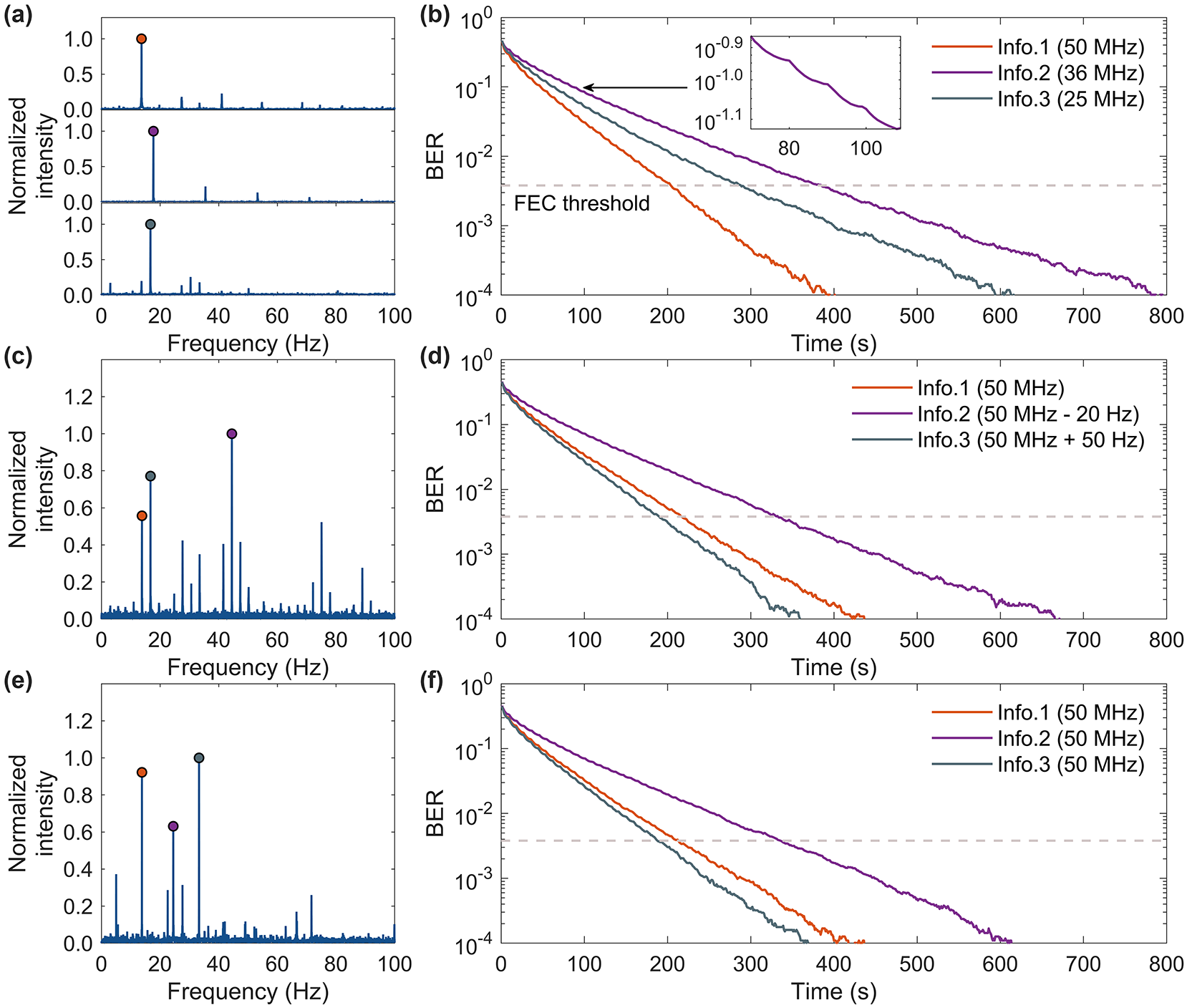}
	\caption{\textbf{Wide frequency adaptability of SPFT scheme under 107 dB channel loss.} 
		\textbf{(a)} The separated frequency domain waveform of three signals when the pulse repetition frequencies among transmitting terminals are significantly distinct. \textbf{(b)} The results of bit error rate (BER) versus correlation time. Different colors represent different transmitting terminals sending the specific information. The average output power of LDs is 0.38 mW. \textbf{(c)} The mixed frequency domain waveform when the pulse repetition frequencies among transmitting terminals differ by 20 to 50 Hz. \textbf{(d)} The corresponding evolution of BER with correlation time. \textbf{(e)} The mixed frequency domain waveform when the pulse repetition frequencies among transmitting terminals are the same. \textbf{(f)} The results of measured BER versus correlation time. FEC: forward error correction threshold. Pushing BER below the FEC threshold is required for error-free information extraction.
	}
	\label{fig3}
\end{figure*}

In the actual application scenario, pulse repetition frequencies of multiple terminals may be random as needed. Thus, the frequency adaptability is one crucial performance, which determines the practicability of our scheme. Here, we set the system channel loss to 107 dB and explore the scheme capability against a wide range of frequencies. We first modulate three 520 nm LDs to $f_1=50$ MHz, $f_2=36$ MHz, and $f_3=25$ MHz, making the pulse repetition frequencies among transmitting terminals are significantly distinct. In this situation, the data process via the scheme is similar to retrieving information when only one transmitting terminal exists. By setting the value of $T_0$ to $1/f_1$, $1/f_2$, and $1/f_3$ respectively, we can extract individual frequency domain waveform belonging to each signal [Fig. \ref{fig3}(a)]. The highest peak in each subgraph represents the linear drift rate between the receiver and the transmitter clock. It can be seen that the extraction of one signal will not be seriously affected by the other signals. In Fig. \ref{fig3}(b), we show the retrieved BER evolution of each terminal, which exhibits a periodic oscillation decreasing over time consistent with the theory described in Section \ref{Methods}. In addition, we can always push BER below the FEC threshold, indicating the high-fidelity information extraction of all terminals.

With the pulse repetition frequencies of the other two transmitters increasing, the frequency difference among transmitting terminals drops to tens of Hertz. In this situation, we can also characterize all effect peaks and draw photons from the sparse single-photon stream to correct terminals by applying the SPFT scheme. Due to the similar pulse frequencies among transmitting terminals, a value set $T_0=1/f_1$ is enough to extract all implicit correlations among photons [Fig. \ref{fig3}(c)]. The BER time evolutions of all terminals shown in Fig. \ref{fig3}(d) match with the results in Fig. \ref{fig3}(b), representing that we can hold the system performance robustness in expansive frequency combination.

Even when the pulse repetition frequencies of all terminals are exactly the same at their local reference, we can still distinguish the slight random frequency drift [Fig. \ref{fig3}(e)] and separate target photons from the sparse photon stream [Fig. \ref{fig3}(f)]. In short, the high resolution about global time correlation features of the scheme enable the precise classification of photons and can synchronously extract multiple weak signals in a wide frequency range. 

\subsection*{B. Ultra-high loss tolerance of SPFT scheme}

\begin{figure*}[tp]
	\centering
	\includegraphics[width=1.5\columnwidth]{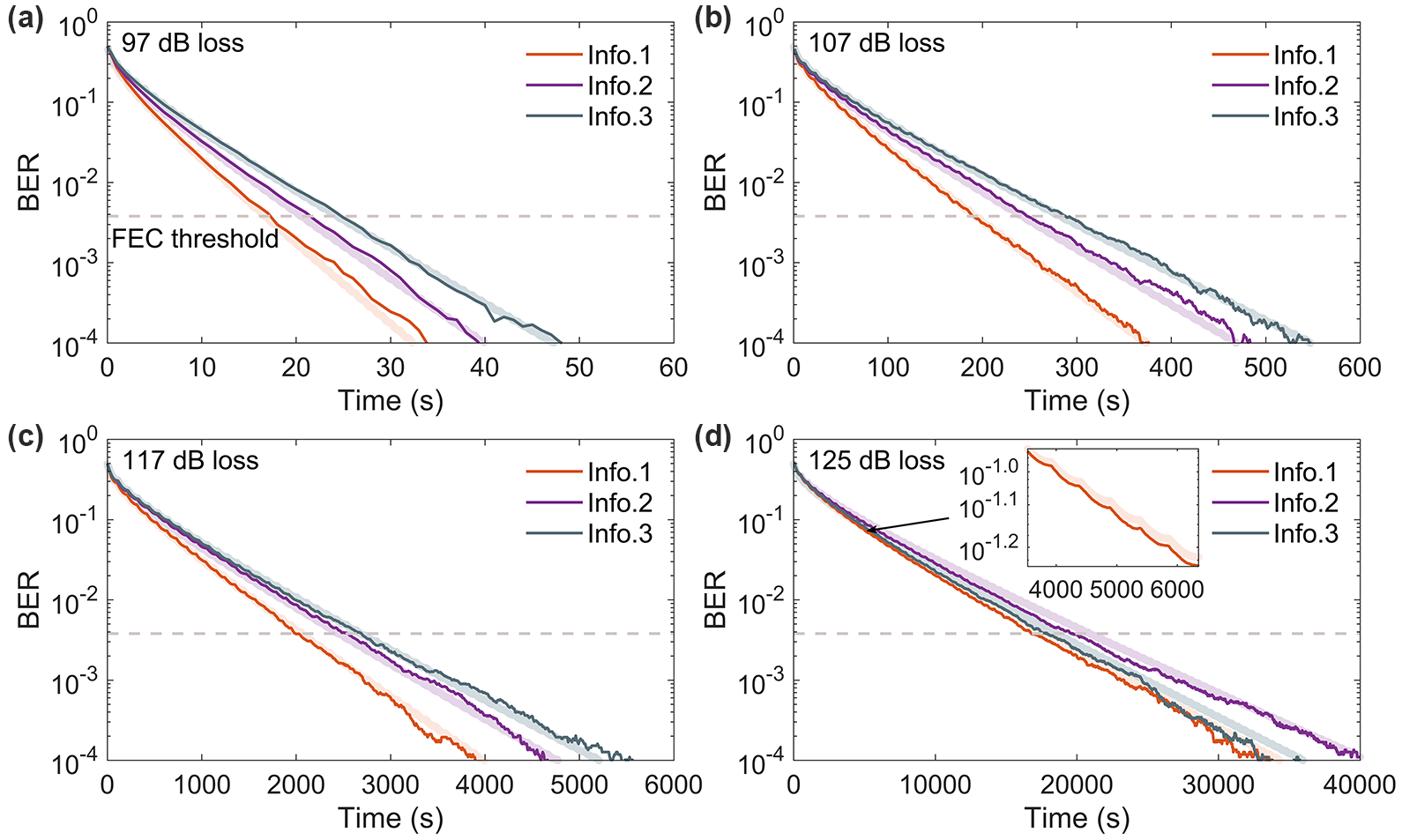}
	\caption{\textbf{Strong loss tolerance of SPFT scheme.}
		\textbf{(a)} The curves of BER versus correlation time under 97 dB channel loss. The dark-colored and light-colored curves represent the experimental results and theoretical expectations, respectively. All transmitting terminals have a pulse repetition frequency of 50 MHz, and the average output power of LDs is 0.31 mW. The signal intensity of each transmitting terminal received is around 80,000 counts per second (cps), corresponding to about $10^{-3}$ photons per pulse survived at the receiving terminal. \textbf{(b)}-\textbf{(d)} The BER time evolution of all signals under 107 dB, 117 dB, and 125 dB channel loss, respectively. As the corresponding survived photon numbers in each pulse are about $10^{-4}$,  $10^{-5}$, and $10^{-6}$. The average output power of LDs is 0.4 mW.
	}
	\label{fig4}
\end{figure*}

Loss tolerance is another crucial performance relating to maximum information transmission distance. Longer transmission distance usually maps to higher channel loss in applications such as optical communication \cite{FSOrelay} and imaging \cite{li2021single}. By manipulating the number of attenuators, we further research the scheme's robustness against various channel losses [Fig. \ref{fig4}]. The output laser power of each transmitting terminal is around 0.4 mW. We first set channel loss to 97 dB, the corresponding BER time evolutions are shown as the dark-colored curves in Fig. \ref{fig4}(a). Compared with the light-colored theoretical BER time evolution in the figure, our experimental results agree well with expectations, representing the precise error control of the scheme against high channel loss (see Section \ref{Methods} for theoretical derivation of BER). 

We gradually increase the channel loss of the system by adding more attenuators. However, due to the global encoding feature, we can always overcome the loss and push the BER below the FEC threshold by extending the detection time and correlating more photons, realizing the high-fidelity extraction of multiple information from the mixed sparse photon stream [Figs. \ref{fig4}(b) and \ref{fig4}(c)]. Finally, we push the channel loss to 125 dB, where only $10^{-6}$ photons per pulse survived at the receiving terminal [Fig. \ref{fig4}(d)]. Although in this extremely sparse photon stream case, our SPFT scheme can still extract target signal photons according to the internal time correlation shared in survived photons.

\subsection*{C. Application of the SPFT scheme in free-space optical communication networks}

\begin{figure*}[bp]
	\centering
	\includegraphics[width=1.8\columnwidth]{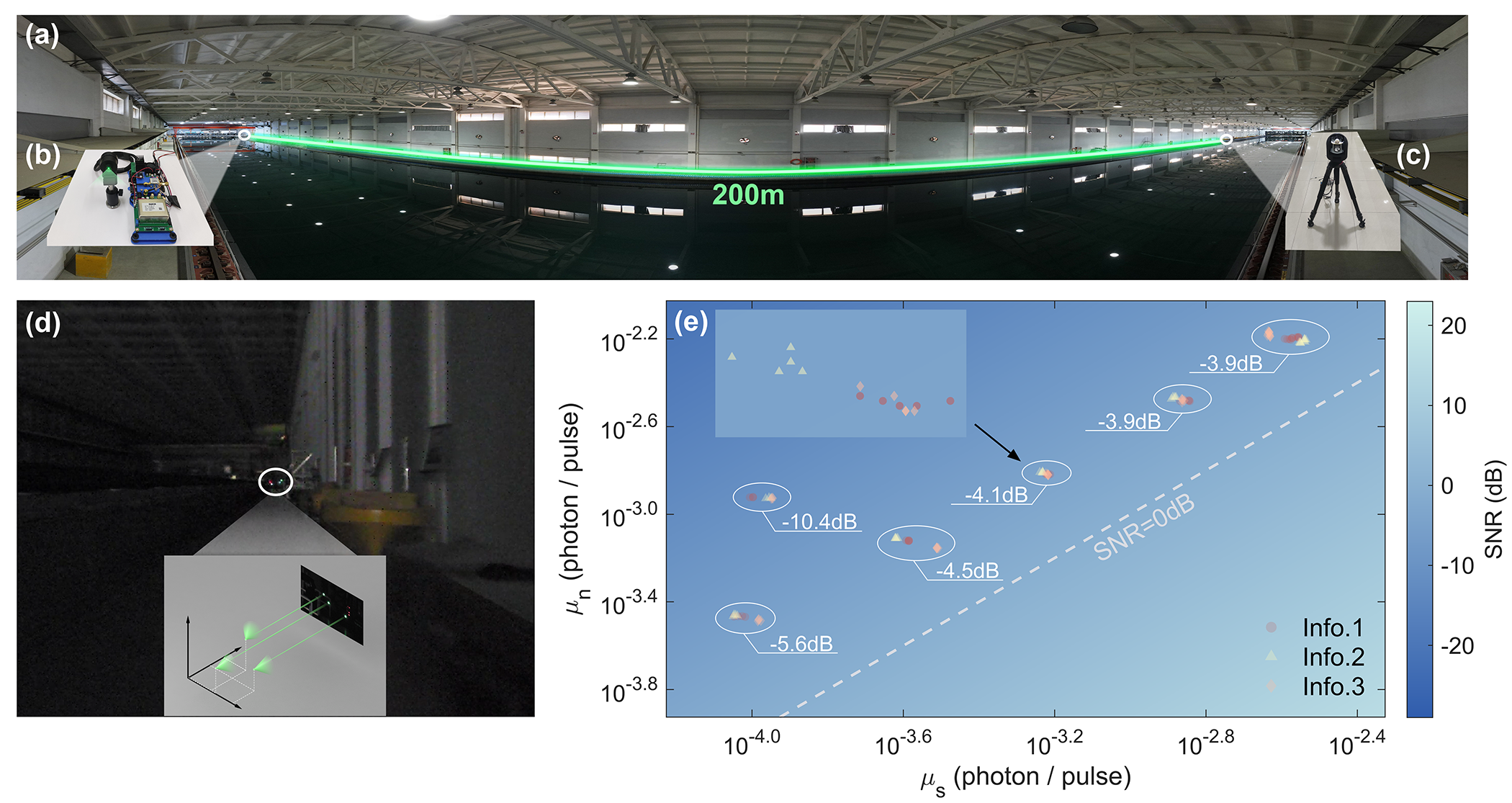}
	\caption{\textbf{Application of SPFT scheme in free-space optical communication networks}
		\textbf{(a)} Panoramagram of the experimental facility. \textbf{(b)} Illustration of a transmitting terminal. A lens group is mounted after the laser to expand the beam divergence angle to $90\,^{\circ}$. The average output power of the three transmitting terminals is 0.36 mW, and the pulse repetition frequencies are 50 MHz. \textbf{(c)} Illustration of the receiving terminal. A wide-angle lens is employed to couple the mixed signals into a multi-mode fiber. The fiber is then connected to a single-pixel single-photon detector. Meanwhile, we use a time-to-digital converter to record the arrival time information of the photon stream. \textbf{(d)} Schematic diagram of the spatial distribution of three transmitting terminals, with a distance of 2-5 m among them. \textbf{(e)} The performance of the system against various SNR environments. Despite the scene where the sparse single-photon signal is embedded in a noise ten times higher, our scheme can still establish multiple communication links only using one detector. The value of $\mu_n$ characterizes the average number of noise photons per pulse, and the value of $\mu_s$ characterizes the signal photons per pulse.
	}
	\label{fig5}
\end{figure*}

The wide frequency adaptability and ultra-high loss tolerance of the scheme allow us to extract multiple time-individual weak signals in various high channel loss systems. Besides the excellent test performance, we demonstrate a practical application of our scheme namely in free-space optical communication networks. A typical communication network consists of massive many-to-one communication units, for instance, the in-orbit optical communication network, which requires a satellite node the ability to decode signals from multiple remote terminals. However, existing weak signal extraction technologies are limited to transmitting single information, resulting in a communication terminal that only can interact with one remote node \cite{Inorbittest, lunarLC}. One solution to expand the multi-channel communication ability is to equip a satellite node with multiple terminals \cite{opticalsatellitenetworks}, which seriously increases the satellite's size and weight.

Here, by introducing the SPFT scheme, we provide an alternative solution to construct a many-to-one communication unit only using one small-sized receiving terminal. We experimentally demonstrate the practicability of the communication unit in a large-scale semi-open facility. By setting three independent transmitting terminals and one receiving terminal on two ends of the facility, we obtain a controllable 200-meter-long free-space channel [Fig. \ref{fig5}(a)]. At each transmitting terminal, to realize the large-spot coverage of multiple receiving terminals, we use a lens group to expand the beam divergence angle to $90\,^{\circ}$ [Fig. \ref{fig5}(b)]. Correspondingly, a wide-angle lens is introduced at receiving terminal to ensure the wide receiving field of view for transmitting terminals [Figs. \ref{fig5}(c) and \ref{fig5}(d)]. The signal intensity detected by receiving terminal ranges from $10^{-4}$ to $10^{-3}$ photons per pulse, according to the pointing direction of terminals. By further switching on the light equipment in the facility, we can increase the ambient noise. Together with the change of signal intensity, we realize the manipulation of SNR in our system (see Supplementary for details). The results in Fig. \ref{fig5}(e) show that despite enduring enormous channel loss and low SNR conditions, our scheme can successfully construct a many-to-one communication unit in various noisy environments. Notably, the strong robustness of the many-to-one communication unit against low SNR is vital owing to the inevitable ambient noise in practical applications. And the noise robustness down to -10.4 dB SNR verifying our scheme's broad application prospect.

\section{Conclusion}

In summary, we propose and experimentally demonstrate a single-photon Fourier transform scheme based on the implicit time correlation shared in the mixed sparse photon stream, enabling the simultaneous extraction of multiple weak signals only using one detector. The experimental results show that our scheme has wide frequency adaptability, which allows us to separate mixed signals with modulation rate differences ranging from several million Hertz to a few Hertz. Even when the pulse repetition frequencies of all terminals are the same, our scheme can utilize the slight random drift between terminals to recover high-fidelity information. Loss tolerance and noise robustness are the other two key factors, dominating the practical application of our scheme. By pushing the loss tolerance up to 125 dB and SNR robustness down to -10.4 dB, we experimentally verify the scheme's reliability against various extreme environments.

Furthermore, we apply the scheme in a practical communication system. By setting three transmitting terminals and one receiving terminal in a large-scale semi-open facility, we construct a basic many-to-one communication unit of free-space optical communication networks. What expands to systems with more transmitting terminals is also achievable. The successful establishment of three communication links indicates that a small-sized SPD is enough to extract multiple information by analyzing the internal correlation among photons. The result is encouraging because our scheme can effectively optimize the size, weight, and complexity of satellite nodes, giving the possibility to construct global-scale in-orbit optical communication networks with complex topology. The feature of separating ultra-weak mixed signals can also apply in the three-dimensional single-photon imaging system. By combining our scheme with multi-source illumination and detection, it is expected to realize a wide-range imaging network with high photon efficiency.

\section{Methods}
\label{Methods}
\subsection*{A. Extraction theory of linear term of clock drift}
\textbf{For only transmitting one signal,} due to the modular arithmetic of photon arrival time $t_a$ being performed relative to pulse period $T_0$: $t_{mod}=t_{a}\ mod\ T_0$, the fluctuation of coincidence delay with time has a certain periodicity. Once the clock drift exceeds $T_0$, the coincidence delay will produce a step change, making it oscillate in a sawtooth shape as a whole (see Supplementary Fig. S2(a)), which can be described by the following formula: 
\begin{equation}
\begin{array}{lc}
\begin{aligned}
x\left(t\right)&=\sum\limits_{n=1}^{+\infty}\frac{T_0 }{\pi n}\left(-1\right)^{n+1}\sin\left(2\pi n\frac{ k_0}{T_0} t+\varphi_0\right)+\frac{T_0}{2},
\end{aligned}
\end{array}
\end{equation}
where $\varphi_0$ is a constant representing the initial phase of sawtooth wave $x\left(t\right)$, $k_0$ is the linear drift rate of the receiver clock relative to the transmitter clock, namely, the drift time per second.

The Fourier expansion of sawtooth wave $x\left(t\right)$ is
\begin{equation}
\begin{array}{lc}
\begin{aligned}
x\left(t\right)&=\sum\limits_{m=-\infty}^{+\infty}X\left(m\frac{k_0}{T_0}\right)\exp{\left({\rm {\mathbf i}} 2\pi n \frac{k_0}{T_0} t\right)}.\\
\end{aligned}
\end{array}
\end{equation}

By using Fourier transform, we get

\begin{equation}
\begin{array}{lc}
\begin{aligned}

X\left(m \frac{k_0}{T_0}\right)=&\frac{1}{T}\int_{0}^{T}x\left(t\right)\exp{\left(-{\rm {\mathbf i}} 2\pi m\frac{ k_0}{T_0} t\right)}\mathrm{d}t\\
=&\frac{1}{T}\int_{0}^{T}\left[\sum\limits_{n=1}^{\infty}\frac{T_0 }{\pi n}\left(-1\right)^{n+1}\sin\left(2\pi n \frac{k_0}{T_0} t\right.\right.\\
&\left.\left.+\varphi_0\right)+\frac{T_0}{2}\right]\exp{\left(-{\rm {\mathbf i}} 2\pi m \frac{k_0}{T_0} t\right)}\mathrm{d}t\\
=&\sum\limits_{n=1}^{\infty}\frac{{\rm {\mathbf i}} T_0 }{2\pi n}\left(-1\right)^{n}\left[\exp{\left({\rm {\mathbf i}}\varphi_0\right)}\delta_{n,m}\right.\\
&\left.-\exp{\left(-{\rm {\mathbf i}}\varphi_0\right)}\delta_{n,-m}\right],\\

\end{aligned}
\end{array}
\end{equation}

where $T$ is the oscillation period of sawtooth wave $x\left(t\right)$. The term $\delta_{n,m}$ is Kronecker delta function
\begin{equation} 
\delta_{n,m}=
\left \{
\begin{aligned}
1,  \text{if} \, \ n=m \\
0,  \text{if} \, \ n\neq m\\
\end{aligned}
\right.
.
\end{equation}
This shows that we can extract the linear term of clock drift by applying Fourier transform to experimental coincidence delay data (see Supplementary Fig. S2(b)).

\textbf{For receiving multiple signals simultaneously,} the coincidence delay is a more complex superposed result of several independent ones (see Supplementary Fig. S2(c)), and its oscillation wave can be theoretically expressed as
\begin{equation}
\begin{array}{lc}
\begin{aligned}
x^{\prime}\left(t\right)&=\sum\limits_{j=1}^{s}\sum\limits_{n=1}^{+\infty}\frac{p_j T_0}{\pi n}\left(-1\right)^{n+1}\sin\left(2\pi n \frac{k_j}{T_0} t+\varphi_j\right)+\frac{T_0}{2},\\
\end{aligned}
\end{array}
\end{equation}
where $j$ indicates different signals and the total number of received signals is $s$. In our demonstrated experiments, we choose $s=3$. $p_j$ means the proportion of $j$-signal in the mixture, which satisfies $\sum\limits_{j=1}^{s}p_j=1$. $\varphi_j$ represents the initial phase of the independent coincidence delay oscillation belonging to $j$-signal. $k_j$ is the linear drift rate of the receiver clock relative to the transmitter clock sending $j$-signal.

The Fourier expansion of sawtooth wave $x^{\prime}\left(t\right)$ is
\begin{equation}
\begin{array}{lc}
\begin{aligned}
x^{\prime}\left(t\right)&=\sum\limits_{m=-\infty}^{+\infty}X^{\prime}\left(m f^{\prime}\right)\exp{\left({\rm {\mathbf i}} 2\pi m f^{\prime} t\right)}.\\
\end{aligned}
\end{array}
\end{equation}

Then applying the Fourier transform, we can extract all implicit periodic features:
\begin{equation}
\begin{array}{lc}
\begin{aligned}
X^{\prime}\left(m f^{\prime}\right)=&\frac{1}{T^{\prime}}\int_{0}^{T^{\prime}}x^{\prime}\left(t\right)\exp{\left(-{\rm {\mathbf i}} 2\pi m f^{\prime} t\right)}\mathrm{d}t\\
=&\frac{1}{T^{\prime}}\int_{0}^{T^{\prime}}\left[\sum\limits_{j=1}^{s}\sum\limits_{n=1}^{+\infty}\frac{p_j T_0}{\pi n}\left(-1\right)^{n+1}\sin\left(2\pi n \frac{k_j}{T_0}  t\right.\right.\\
&\left.\left.+\varphi_j\right)+\frac{T_0}{2}\right]\exp{\left(-{\rm {\mathbf i}} 2\pi m f^{\prime} t\right)}\mathrm{d}t\\
=&\sum\limits_{j=1}^{s}\sum\limits_{n=1}^{+\infty}\frac{{\rm {\mathbf i}} p_j T_0}{2\pi n}\left(-1\right)^{n}\left[\exp{\left({\rm {\mathbf i}} \varphi_j\right)}\delta_{n \frac{k_j}{T_0},m f^{\prime}}\right.\\
&\left.-\exp{\left(-{\rm {\mathbf i}} \varphi_j\right)}\delta_{n \frac{k_j}{T_0},-m f^{\prime}}\right],\\
\end{aligned}
\end{array}
\end{equation}
where $T^{\prime}=1/f^{\prime}$, $f^{\prime}$ represents the fundamental frequency of Fourier expansion of the composite sawtooth wave $x^{\prime}\left(t\right)$. In general, three primary characteristic peaks in the frequency domain waveform correspond to the linear clock drift rates of three signals, the rest are their higher frequency components describing the morphological features of each oscillating waveform respectively or caused by ambient noise (see Supplementary Fig. S2(d)).

\subsection*{B. Theoretical derivation of BER}
After experiencing huge attenuation, the signal detected by the receiving terminal is a weak coherent state. Thus, the photon number distribution of each signal pulse follows a Poisson distribution \cite{yan2021PICOC, Decoy2005, hu2007poisson}. The noise in each pulse includes photons of the other two signals and the randomly arriving ambient noise photons \cite{noisePoisson}, which also obey the Poisson distribution. Therefore, the photon statistics of 0-bit and 1-bit can be expressed as:

\begin{equation}
\begin{array}{lc}
\begin{aligned}
\hspace{-1.5mm}
P_{0,i}\left(t,n_p\right)=&a_{0,i}\frac{\left(\sum\limits_{j=1,j\neq i}^{s}\mu_{j} t\!+\!\nu t\right)^{\!n_p}}{n_p!}\exp{\left[\!-\!\left(\sum\limits_{j=1,j\neq i}^{s}\mu_{j} t\!+\!\nu t\right)\right]}\\
\hspace{-1.5mm}
P_{1,i}\left(t,n_p\right)=&a_{1,i}\frac{\left(\sum\limits_{j=1}^{s}\mu_{j} t\!+\!\nu t\right)^{n_p}}{n_p!}\exp{\left[\!-\!\left(\sum\limits_{j=1}^{s}\mu_{j} t\!+\!\nu t\right)\right]}\\
\end{aligned}
\end{array}
,
\end{equation}
where $P_{0,i}$ and $P_{1,i}$ depict the statistical probabilities of 0-bit and 1-bit, subscript $i$, $j$ indicate different signals and the total number of received signals is $s$, here, $s=3$. $t$ is the correlation time as $n_p$ denotes the photon mumber. $a_{0,i}$ and $a_{1,i}$ express the ratio of 0-bit and 1-bit belonging to $i$-signal, $\mu_{j}$ and $\nu$ represent the average photon number of $j$-signal and ambient noise per bit in a second duration, respectively.

Then the theoretical expectation of BER time evolution can be expressed as
\begin{equation}
\begin{array}{lc}
\begin{aligned}
BER_i\left(t\right)=\sum\limits_{n_p=N_{p,i}\left(t\right)}^{+\infty}P_{0,i}\left(t,n_p\right)+\sum\limits_{n_p=0}^{N_{p,i}\left(t\right)-1}P_{1,i}\left(t,n_p\right),&\\
i=1,2,…,s.&\\
\end{aligned}
\end{array}
\end{equation}
where $N_{p,i}\left(t\right)$ represents the photon number threshold to judge 0-bit or 1-bit according to the photon number, which allows us to minimize the BER. By calculating $P_{0,i}\left(t,n_p\right)=P_{1,i}\left(t,n_p\right)$, we can obtain the expression of $N_{p,i}\left(t\right)$.

\begin{equation}
\begin{array}{lc}
\begin{aligned}
N_{p,i}\left(t\right)=\left\lceil \frac{\mu_{i} t+\ln\left(a_{0,i}/a_{1,i}\right)}{\ln\left[\left(\sum\limits_{j=1}^{s}\mu_{j}+\nu\right)\bigg/\left(\sum\limits_{j=1,j\neq i}^{s}\mu_{j}+\nu\right)\right]} \right\rceil.\\
\end{aligned}
\end{array}
\end{equation}

The time evolution of $N_{p,i}\left(t\right)$ is a step function, which causes the BER curve exhibiting a periodic fluctuation with time.

\section*{Funding.}
National Key R\&D Program of China (Grant No. 2024YFA1409300); National Natural Science Foundation of China (NSFC)
(Grants Nos. 62235012, 12304342, 12574549, 12574542); Quantum Science and Technology-National Science and Technology Major Project (Grant Nos. 2021ZD0301500 and 2021ZD0300700); Science and Technology Commission of Shanghai Municipality (STCSM) (Grant Nos. 2019SHZDZX01,
24ZR1438700, 24ZR1430700 and 24LZ1401500); Startup Fund for Young Faculty at SJTU (SFYF at SJTU) (Grant Nos. 24X010502876 and 24X010500170); Frontier Technologies R\&D Program of Jiangsu (Grant No. SBF20250000094)

\section*{Acknowledgments.}
X.-M.J. acknowledges additional support from a Shanghai talent program.

\section*{Disclosures.}
The authors declare no conflicts of interest.

\section*{Data availability.}
Data underlying the results presented in this paper are not publicly available at this time but may be obtained from the authors upon reasonable request.

\section*{Supplementary document.}
See Supplementary Materials for supporting content.

\end{document}